\documentclass[11pt]{article}

\usepackage[margin=0.8in]{geometry}
\usepackage{amsmath, amssymb, amsthm}
\usepackage{bm}
\usepackage{mathtools}
\usepackage{graphicx}
\usepackage{booktabs}
\usepackage{multirow}
\usepackage{array}
\usepackage{enumitem}
\usepackage{caption}
\usepackage{subcaption}
\usepackage{float}
\usepackage{setspace}
\usepackage{xcolor}
\usepackage{hyperref}
\usepackage{comment}

\usepackage{booktabs}
\usepackage[table]{xcolor}

\graphicspath{{Fig/}}
\usepackage{bm}
\usepackage[numbers,sort&compress]{natbib}

\hypersetup{
    colorlinks = true,
    linkcolor = blue,
    citecolor = blue,
    urlcolor = blue
}

\theoremstyle{thmstyleone}%
\newtheorem{theorem}{Theorem}
\newtheorem{proposition}[theorem]{Proposition}%
\theoremstyle{thmstyletwo}%
\newtheorem{remark}{Remark}%
\theoremstyle{thmstylethree}%

{}
\newcommand{\lf}{\left}
\newcommand{\rt}{\right}

\newcommand{\nn}{\nonumber}
\newcommand{\bY}{\mathbf{Y}}
\newcommand{\bW}{\mathbf{W}}
\newcommand{\bP}{\mathbf{P}}
\newcommand{\bX}{\mathbf{X}}

\newcommand{\bZ}{\mathbf{Z}}

\title{
Testing Procedures for Strict Pleiotropy in Genetic Association Studies
}

\author{
Eva Biswas$^{1}$ \and
Nilanjan Chatterjee$^{1,2}$ \and
Zheyu Wang$^{1}$
}

\date{}

\begin{document}

\maketitle


\begin{center}
\small
$^{1}$Department of Oncology, Johns Hopkins University, School of Medicine, Baltimore, MD\\
$^{2}$Department of Biostatistics, Johns Hopkins Bloomberg School of Public Health, Baltimore, MD
\end{center}


\abstract{Pleiotropy, the phenomenon in which a single genetic variant influences multiple traits, is a fundamental feature of genome biology and an important consideration in statistical genetics. Many statistical methods exist that leverage pleiotropy to increase the power of association tests, but they are often designed to identify SNPs which are associated with at least one of the underlying traits. Only a limited number of methods exist for identification of ``strict pleiotropy", i.e. those associated with more than one trait, but these methods are either restricted to a small number of traits or become computationally challenging as the number of traits increases. In this paper, we develop two testing procedures to identify SNPs under strict pleiotropy and determine the corresponding minimum number of associated traits. Both methods rely only on z-scores available from genome-wide association studies (GWAS) and are applicable to independent as well as correlated traits. The first procedure is frequentist in nature and is designed to control the family-wise-error rate. The second procedure is based on the local false discovery rate and provides improved power while maintaining FDR control. We conduct extensive simulation studies to compare the performance of the proposed methods with alternative methods for testing strict pleiotropy. Finally, we demonstrate an application of the proposed methods to Alzheimer’s disease-related biomarkers for the goal of distinguishing SNPs that may represent biomarker-specific variations from those that are related to disease-relevant biology.
}

\vspace{1cm}


\maketitle

\section{Introduction}
Pleiotropy, the phenomenon where a single genetic variant influences multiple traits \cite{solovieff2013pleiotropy,stearns2010one}, is a fundamental property of genome biology and a key concept in statistical genetics. Genome-wide association studies (GWAS) have provided evidence of widespread pleiotropy across a broad range of human traits \cite{atwell2010genome,pickrell2016detection,qi2024genome, visscher2016plethora,watanabe2019global}, demonstrating that genetic architecture is highly interconnected.  Researchers have leveraged pleiotropy to improve the power of SNP discovery in GWAS.  Assuming that the traits are associated through pleiotropy, multivariate and multi-trait GWAS methods have been developed to jointly test associations between genetic variants and multiple related phenotypes, thereby improving power to detect SNPs associated with the traits \cite{yang2016efficient,yang2017identifying,baker2009multivariate,heid2017multitrait,bhattacharjee2012subset,ray2018methods}. However, these approaches primarily test whether a SNP is associated with \textit{at least one trait} within a phenotype set, rather than explicitly testing for multi-trait per se.



Several methods \citep{zhang2014estimating,ray2020powerful,ray2021pleiotropy,park2025robust} have been developed to test for ``strict” multi-trait association under suitable composite null hypotheses. However, these methods are difficult to generalize beyond two traits and fail to account for genetic correlations among traits and for correlations induced by GWAS sample overlap.  \cite{schaid2016statistical} introduced a sequential testing framework for pleiotropy across more than two traits, though their method requires individual-level data, which limits its applicability, as in practice, we may not have access to individual-level data. In addition to traditional multivariate testing frameworks, empirical Bayes approaches based on the local false discovery rate (lfdr) \citep{efron2002empirical,efron2008microarrays,sun2007oracle} have become useful tools for testing genetic effects across multiple traits. Recently, \citet{sun2025testing} proposed the conditionally symmetric multidimensional Gaussian mixture model (csmGmm) for testing strict pleiotropic association while controlling the false discovery rate (FDR), but this method faces substantial computational challenges when the number of traits is large.


 Testing for strict pleiotropy becomes highly relevant when investigating the genetic association of latent health conditions that manifest through multiple biomarkers. For example, the early stage of Alzheimer’s disease (AD), when cognitive symptoms are not yet developed, is characterized by alterations in multiple biomarkers, including amyloid beta, phosphorylated tau (p-tau), neurogranin, synaptosomal-associated protein 25 (SNAP-25), and neurofilament light (NFL) \cite{hampel2018blood, hansson2021biomarkers}. Likewise, inflammation underlies various complex diseases, including cardiovascular, autoimmune disorders, and metabolic disorders, and is commonly assessed using biomarkers such as C-Reactive Protein (CRP), high-sensitivity CRP, Erythrocyte Sedimentation Rate (ESR), immunoglobulin E (IgE), and tumor necrosis factor (TNF) \cite{musaad2007biomarkers, remmelzwaal2022inflammation}. In such cases, genetic variants associated with individual biomarkers may simply reflect biomarker-specific biological processes. But variants that co-regulate multiple biomarkers simultaneously are more likely to provide insight into the biological mechanisms of underlying latent conditions.  Further,  identifying such pleiotropic variants can also lead to a proper genetic ``instrument" for testing for causal effects of latent exposures like inflammation on health conditions \cite{jin2024mendelian}. When the number of biomarkers is relatively large,  it may be, however,  too restrictive to identify SNPs which are associated with all traits, and a more suitable formulation for the alternative hypothesis would be to identify SNPs which show a certain ``degree" of pleiotropy, i.e. they are associated with a minimum number, say, $r$, of the $K$ traits

In this study, we revisit the methodology for testing for strict pleiotropy and make several novel contributions. We propose two testing methodologies to identify pleiotropic SNPs by using the z-scores readily available from GWAS studies. The first method uses a minimum of z-scores to test the alternative that a SNP affects all traits, and we demonstrate that this simplified test statistic is asymptotically equivalent to the likelihood-ratio test requiring individual-level data. For the general alternative that a SNP affects at least $r$ of $K$ traits, we develop a sequential test based on z-score quantiles while controlling the family-wise error rate (FWER). The advantages of the test are that it is simple, it does not involve any complex estimation method, and hence it is computationally efficient to implement, even when the number of traits is large. Moreover, its performance is robust to the dependence structure among the traits.


Within the same framework, we further propose a FDR control method for testing the pleiotropy architecture across SNPs.
The test accounts for the correlation among observed traits arising from overlapping samples, as well as the correlation of SNP effects across traits due to genetic correlation. While csmGmm models the mean structure of z-scores, our method models the variability of z-scores across SNPs by introducing fewer parameters to reduce computational burden for a large number of biomarkers. We also provide a step-by-step estimation procedure that facilitates the implementation of the proposed test and improves computational efficiency. We conduct extensive simulation studies to compare the performance of these novel methods with that of existing methods. We illustrate the application of these two methods through pleiotropic analysis of established Alzheimer's disease biomarkers using GWAS summary statistics.

\section{Statistical Framework and Methods}\label{Sec:Methods}
Consider $K$ traits and suppose the objective is to identify SNPs that influence all or a subset or none of these traits simultaneously. For each trait $k=1,..,K$, let ${\bY}_k = \{Y_{k1}, \ldots, Y_{kn_k}\}$ denote the observed values of the $k^{\text{th}}$ trait across $n_k$ individuals. Suppose ${\mathbf{S}}_{kj}=\{{S}_{kj1},...,{S}_{kjn_k}\}$ is the vector of $j^{\text{th}}$ genetic variant for all $j = 1, \ldots, p$ in the dataset for $k^{th}$ biomarker. It is important to note that the datasets corresponding to different traits may partially or completely overlap in individuals.

In GWAS, a quantitative trait is typically modeled using a linear regression by assuming
\begin{align}\label{eq: model eq}
  \bY_k=\beta_{jk0} + \beta_{jk} {\mathbf{S}}_{kj} +  {\mathbf{Z}}_k^\top\bm{\gamma}_{jk}+\bm{\epsilon}_{jk}
\end{align}
where $\beta_{jk}$ represents the effect size (slope) of $j^{th}$ genetic variant. The vector $\bm{Z}_k$ includes other covariates, such as age, sex, and genetic principal components, and $\epsilon_{jk}$ is a vector of random errors. When the trait of interest is binary (e.g., disease status), a logistic regression model is used instead.

 The estimated slope $\widehat\beta_{jk}$, its standard error $\mathrm{se}(\widehat{\beta}_{jk})$ and the Wald test statistic (or z-score ) $W_{jk}=\frac{\widehat{\beta}_{jk}}{\mathrm{se}(\widehat{\beta}_{jk})}$ are publicly available from GWAS summary statistic. Our two sequential testing procedures use these Wald statistics as their primary inputs.

\subsection{Formulation of the hypothesis}\label{sec:statement}
Consider the hypothesis where the goal is to detect variants that affect at least $r$ out of $K$ traits. The test is formulated as:
\begin{align}\label{Hypothesis Statement 2}
\textbf{Test:} \quad H_0^r: &\beta_{jk} = 0 \text{ for all } k \in C^c \subset \{1,\ldots,K\}, \text{ for any set } |C| < r \nn\\
\text{vs} \quad H_a^r:& \beta_{jk} \ne 0 \text{ for all } k \in C, \text{ for some set $C$ with $|C|\ge r$}
\end{align}
Under
$H_0^r$, the SNP influence atmost $r-1$ traits, i.e., no group of
$r$ or more traits simultaneously show nonzero effects. Equivalently, the alternative hypothesis $H_a^r$
asserts that there exists a subset $C$ of at least
$r$ traits for which the variant has a nonzero effect.


\subsection{The minimum-test and a connection with likelihood-ratio} \label{sec:min test}

We first test whether a SNP affects all traits simultaneously to assess ``strict pleiotropy''. We propose a simple test using the Wald statistic, which is readily available, and show that the test is asymptotically equivalent to the likelihood ratio test, which requires individual-level data.

First, we fix a variant $j$ and discuss the methodology to test the hypothesis \eqref{Hypothesis Statement 2} for $r=K$, where the alternative asserts that the variant influences all traits. We begin by considering the individual hypotheses for each trait $k=1,\ldots,K$ as
\begin{align}\label{eq: sub Hypothesis}
 H_{0k}: \beta_{jk}= 0 \mbox{ vs } H_{ak}: \beta_{jk} \ne 0.
\end{align}
The rejection of hypothesis $H_{0k}$ depends on how large the statistic $|W_{jk}|$ is.
 The global null hypothesis $H_0^K = \bigcup\limits_{k=1}^K H_{0k}$, and $H_a=\bigcap\limits_{k=1}^K H_{ak}$. We reject the global null only if all individual nulls $H_{0k}$ are rejected, i.e., all the statistics $W_{jk}$ exceed the corresponding threshold. Therefore, an intuitive and simple test statistic for assessing pleiotropy as the minimum absolute Wald statistic across all traits:
    \begin{equation}\label{eq: Likelihood Test}
    T_{N,j,K} = \min_{k=1,...,K}|W_{jk}|.
  \end{equation}
    where $N = \{n_1,...,n_k\}$. There is no simple form of the asymptotic distribution of the test statistic, particularly under dependence. By applying the intersection-union test (\citealp{casella2024statistical})
    we obtain the rejection region for a significance level $\alpha$, as stated in the following proposition.
\begin{proposition}\label{res: Wald Test}
For $\alpha\in(0,1)$, let $c_\alpha=\Phi^{-1}(1-\alpha/2)$, where $\Phi$ denotes the cumulative distribution function of the standard normal distribution. Then $
\lim_{ N\to\infty}\sup_{\boldsymbol{\theta}\in\Theta_0}
P_{\boldsymbol{\theta}}\left(T_{N,j,K}>c_\alpha\right)=\alpha,$
where $\boldsymbol{\theta}=(\beta_{j1},\ldots,\beta_{jK})^\top$ and $\Theta_0$ denotes the parameter space under $H_0^K$.
\end{proposition}

   The rejection region of each $W_{jk}$ corresponding to $H_{0k}$ is also $R_{\alpha}=\{z:|z|>c_\alpha\}$. Therefore, $H_0^K$ will be rejected if all the individual $W_{jk}$ fall in $R_{\alpha}$.
To perform the above test, we do not need access to individual-level data; we only need the estimated effect size and its standard error, which are available from the GWAS summary statistics. Also, the above test is not affected by correlation among the GWAS data sets.

Next we consider the likelihood-ratio test for the composite null hypothesis $H_{0}^k$, assuming individual-level data are available from $N$ individuals and the traits follow a multivariate Gaussian distribution. Also assume $\bX$ is the full design matrix combining all covariates, and $\bX_r$ is the reduced design matrix obtained by removing the column corresponding to the $j^{\text{th}}$ genetic variant. Then the following proposition presents the test statistic and the corresponding rejection region to perform the hypothesis test $H_0^K$ (see derivation in Sec
\begin{proposition}
The likelihood ratio test is given by
\begin{align}\label{eq:LRT}
    l_N
    = \min_{k = 1, \ldots, K}
      {N}
      \log \left|
      1 + \frac{1}{N}
      \bY_k^\top (\bP_\bX - \bP_{\bX_r}) \bY_k \,
      (\widehat{\Sigma}^{-1})_{kk}
      \right|.
\end{align}
where  $\bP_\bX$ and $\bP_{\bX_r}$ are the projection matrices onto the column spaces of $\bX$ and $\bX_r$, respectively, $\widehat{\Sigma} = \tfrac{1}{n} \bY^\top (\mathbf{I}_N - \bP_\bX) \bY$, with $\mathbf{I}_N$ being the $N\times N$ identity matrix, and  $(\widehat{\Sigma}^{-1})_{kk}$ denotes the $k^{th}$ diagonal element of $\Sigma^{-1}$.\\
Also, for level $\alpha$, $\lim\limits_{n\to \infty}\sup_{{\theta}\in \Theta_0}P_{\theta}(l_n \in  R_{\alpha}')=\alpha$, where $\theta$ is the vector of all the parameters needed to specify the full model for $\bY$, and $R_{\alpha}' = \{z:z>F^{-1}(1-\alpha)\}$ with $F$ being the cumulative distribution function of the $\chi^2(1)$ distribution.
\end{proposition}

See the Supplements for the derivation of the test statistics and the proof of the asymptotic distribution.
\begin{remark}\label{res: lik ratio Test}
The representation of the LRT in the above form suggests the optimality of minimum-test-statistics for testing the composite null hypothesis of interest. In particular, for any individual hypothesis $H_{0k}$, we know that the Wald statistic is asymptotically equivalent to the LRT. In the current setting, since the global null hypothesis is $H_{0}^{K} = \bigcup_{k=1}^K H_{0k}$, the minimum of the individual Wald statistics $T_{N,j,K}$ lies in the rejection region $R_{\alpha}$ if and only if each individual Wald statistic $W_{j,k}$ falls in its respective rejection region. Equivalently, the individual LRT statistics for testing $H_{0k}$ fall in the rejection region $R_{\alpha,k}' = \{ z : z > F^{-1}(1 - \alpha) \} = R_{\alpha}'$ for all $k = 1, \ldots, K$.
Therefore, $T_{N,j,K}$ has an asymptotically equivalent rejection region to the likelihood ratio test for assessing $H_{0}^K$ versus $H_{a}^K$.
\end{remark}

\subsection{A Sequential Quantile-based Test (SQT)}
Given  the optimality property of the min-test for testing against $H_{0}^{K}$, in this section, we will consider a quantile-based test for testing $H_{0}^{r}$ when $r<K$. Further, we note that in GWAS, typically, Bonferroni correction is applied for the number of SNPs tested to maintain FWER. Here, $FWER = P_{H_0^r}(\text{at least one SNP falsely rejecting $H_0^r$})$. In other words, the test for each SNPs is performed at a fixed threshold $\alpha^{*}=\alpha/P$, where $P$ denotes the total number of SNPs. However, in the context of testing pleiotropic effects for SNPs across multiple traits, the use of a fixed threshold across SNPs is not efficient, and we propose a sequential version of the test that can adaptively adjust the significance threshold level at each stage depending on the number of SNPs that passed the previous stage. The algorithm is described step-by-step below.

\textbf{Step 1:} Extract the $p$-values of the common variants from the $K$
GWAS studies after harmonization and create a $p_0\times K$ matrix
$\bP_0$, where $p_0$ is the number of common variants. Also retain the
variant ID corresponding to each row of the matrix.

\textbf{Step 2:} For each SNP $j$, sort the $K$ $p$-values in ascending
order and denote them by
\[
p_{(1),j} \leq p_{(2),j} \leq \cdots \leq p_{(K),j}.
\]

\textbf{Step 3:} Start with $\bP=\bP_0$ and $r=1$. We apply the following
iterative procedure, which adaptively filters variants using
multiple-testing-adjusted thresholds:

\begin{enumerate}
    \item[i.] For $r=1,\ldots,K$, let $q$ denote the number of SNPs
    remaining in $\bP$ and calculate
    \[
    c_r=\frac{\alpha}{q(K-r+1)}.
    \]

    \item[ii.] Retain the SNPs $j$ satisfying
    \[
    p_{(r),j}<c_r,
    \]
    and update $\bP$ by keeping only these SNPs. Record the corresponding
    SNP IDs as variants identified as being associated with at least $r$
    traits.
\end{enumerate}

While running the procedure, if no SNPs survive at iteration $r=r_0$, this indicates that there are no SNPs associated with more than $r_0-1$ biomarkers. Furthermore, if the goal is to identify variants influencing at least $r$ biomarkers, the above algorithm is applied sequentially from $k=1$ through $r$.
 The quantity $K-r+1$ in $\alpha/(K-r+1)q$ controls the type one error for each variant to test $H_0^r$, while the quantity $q$ maintains the FWER at $\alpha$ for testing across $q$ many variants selected in the previous step, which leads to the following proposition.
\begin{proposition}\label{prop: sqt}
     For SQT with nominal level $\alpha$, the FWER is maintained at $\alpha$ at each iteration $r$ corresponding to hypothesis $H_0^r$ asymptotically.
 \end{proposition}

\section{Empirical-Bayes method and FDR control}\label{lfdr}
Thus far, we have developed a testing procedure that controls the FWER. In this section, we present an alternative approach for testing \eqref{Hypothesis Statement 2} that controls the FDR rather than the FWER, with the aim of improving statistical power. The proposed method models the z-scores using a mixture of symmetric Gaussian distributions.



We have $W_{jk}\sim N(0,1)$, if there is no true effect of SNP $j$ on trait $k$, i.e., $\beta_{jk}=0$. Otherwise, the score $W_{jk}$ increases with sample size, as it is computed without centering. So we assume that the score $W_{jk}$ follows $N(0,\sigma_k^2)$ for some large variance $\sigma_k^2$ if $\beta_{jk}\ne 0$, and the marginal distribution can be written as
\begin{align}\label{eq: W dist}
    W_{jk} \sim (1-p_k) N(0,1)+p_k N(0,\sigma_k^2)
\end{align}
with $\sigma_k^2>1$, where $p_k$ corresponds to the probability of a randomly selected SNP $j$ being associated with trait  $k$.
Define $\bZ_j = \{Z_{j1},...,Z_{jK}\}$ as the vector of $1/0$ indicator variables that indicate whether the SNP $j$ is associated with the $k$-th trait across $k=1,\ldots,K$ and $|\bZ_j|=\sum_{k=1}^K Z_{jk}$, denoting the number of corresponding non-zero effects.  Further, we observe that z-scores across a pair of biomarkers $(k,k')$ can be correlated due to the correlation ($\rho_{kk'}^{b}$) in true effect sizes for the SNPs across the biomarkers and due to the correlation in estimation error ($\rho_{kk'}^{e}$) that arises when there is sample overlap in the underlying GWAS. In the appendix, we show that under the proposed modeling framework, one can express $\mathrm{Cov}(W_{jk},W_{jk'}) = \rho_{kk'}^e$ for any SNP $j$ that is associated with none or only one of the traits, and $\mathrm{Cov}(W_{jk},W_{jk'}) =\sqrt{(\sigma_k^2-1)(\sigma_{k'}^2-1)}\rho_{kk'}^b +\rho_{kk'}^e$ for any SNP $j$ that is associated with both traits. Now, if we denote $\mathbf{W}_j=(W_{j1},\ldots,W_{jK})^{T}$ as the vector of observed z-scores associated with the $j$-th SNP, then the local false discovery rate under the composite null hypothesis $H_0^{r}$ can be written as
\begin{align}\label{eq:ldfr}
    lfdr(\mathbf{W}_{j}) :&= P(H_0^r | \mathbf{W}_{j})=\frac{\sum_{\mathbf{z}:|\mathbf{z}|\le r-1} \pi_{\mathbf{z}}\phi_{K}(\mathbf{W}_{j},\Sigma(\mathbf{z}))}{\sum_{\mathbf{z}}\pi_{\mathbf{z}}\phi_{K}(\mathbf{W}_{j},\Sigma(\mathbf{z}))}
\end{align}
where $\pi_{\mathbf{z}}$ denotes the probability of the configuration $\bZ=\mathbf{z}$ and $\phi(\mathbf{W}_j,\Sigma(\mathbf{z}))$ denotes the $K$-variate normal density under mean zero and variance-covariance matrix $\Sigma$, which depends on the configuration $\mathbf{z}$, evaluated at $\mathbf{W}_j$. Evaluation of lfdr in the above setting requires first the estimation of the unknown parameters $\pi_{\cdot}$, i.e the probability distribution of $Z$, and the underlying variance-covariance parameters indexed by $\sigma_k^2$, $\rho_{k,k'}^{b}$ and $\rho_{k,k^{'}}^{e}$, $k,k^{'}=1,\ldots,K$. We optimize using the EM algorithm, accounting for the mixture-likelihood structure induced by the latent factors $\mathbf{Z}$. The performance of the test depends on $\widehat{\pi}_z$. For instance, suppose $\widehat\pi_{\mathbf z_1}\gg\widehat{\pi}_{\mathbf z_2}$, where $\mathbf z_1$ corresponds to a larger subset of traits being associated with a SNP than $\mathbf z_2$. In this case, even when the evidence for association for an individual SNP corresponding to the configuration $z_2$ to be higher than $z_1$, i.e. $\phi_K(\bW,\Sigma(\mathbf{z}_2))>\phi_K(\bW,\Sigma(\mathbf{z}_1))$,  one may have $\widehat{\pi}_{\mathbf{z}_1}\phi_K(\bW,\widehat\Sigma(\mathbf{z}_1))>\widehat{\pi}_{\mathbf{z}_2}\phi_K(\bW,\widehat\Sigma(\mathbf{z}_2))$.
Therefore, to avoid parameter estimates that correspond to substantially higher probabilities for configurations in $Z$ with more non-zero values than with fewer non-zero values, we propose optimizing the penalized complete log-likelihood of the following form using the standard EM algorithm-

\begin{align}\label{eq:mixlik}
l_{\lambda}(\pi,\Sigma) =\sum_{j=1}^p\sum_{\mathbf{z}\in\{0,1\}^K}\{I_j(\mathbf{z})\log \pi_{\mathbf{z}} + I_j(\mathbf{z}) \log \phi_K(\mathbf{W}_{j};\Sigma(\mathbf{z}))\} - \lambda\lf(\sum\limits_{\substack{\mathbf{z}_1,\mathbf{z}_2:|\mathbf{z}_1|-|\mathbf{z}_2|=1\\ \mathbf{z_1}^\top\mathbf{z_2}=|\mathbf{z}_2|}}\frac{\pi_{\mathbf{z}_1}}{\pi_{\mathbf{z}_{2}}}\rt),
\end{align}
where $I_j(\mathbf{z})$ is 1 if $\bZ_j=\mathbf{z}$, otherwise 0,  the sum inside the penalty term corresponds to all pairs of $\mathbf{z}_1$ and $\mathbf{z}_2$ that differ in their value for only one biomarker, with $\mathbf{z}$ denoting the configuration with more associated biomarkers, and the ratio $\pi_{\mathbf{z}_1}/\pi_{\mathbf{z}_2}$
corresponding to the odds of a SNP being associated with the biomarker given that the other configurations of association are fixed. For example, if there are three biomarkers, the penalty term is given by $\lambda(\frac{\pi_{100}}{\pi_{000}}+\frac{\pi_{010}}{\pi_{000}}+\frac{\pi_{001}}{\pi_{000}}+\frac{\pi_{110}}{\pi_{100}}+\frac{\pi_{110}}{\pi_{010}}+\frac{\pi_{011}}{\pi_{010}}+\frac{\pi_{011}}{\pi_{001}}+\frac{\pi_{101}}{\pi_{100}}+\frac{\pi_{101}}{\pi_{001}}+\frac{\pi_{111}}{\pi_{110}}+\frac{\pi_{111}}{\pi_{011}}+\frac{\pi_{111}}{\pi_{101}})$.
The parameter $\lambda$ softly puts restriction such that $\pi_{\mathbf{z}_1}$ is not larger than $\pi_{\mathbf{z}_2}$.

It is important to choose $\lambda$ in a data-driven manner. For a given SNP, let
$
\mathbf z_1
=
\arg\max_{\mathbf z}
\phi_K(\mathbf W,\widehat\Sigma_{\mathbf z})
$. If $z_1$ corresponds to a rare event, then it might happen that
$
\mathbf z_2
=
\arg\max_{\mathbf z}
\widehat\pi_{\mathbf z}\phi_K(\mathbf W,\widehat\Sigma_{\mathbf z})
$
involving more associated traits than $\mathbf z_1$ and $\widehat\pi_{\mathbf z_2}\gg \widehat\pi_{\mathbf z_1}$. Therefore, multiplication by the configuration probabilities can suppress the contribution of $\mathbf z_1$ in the mixture distribution and hence favor the configuration $\mathbf z_2$. We refer to SNPs for which this occurs as discrepant SNPs.
For such SNPs, this discrepancy can lead to a relatively small estimated mixture likelihood. For $\lambda=0$, we therefore aggregate the log-mixture density of the discrepant SNPs, as given by
\[
L_0^{\mathrm{dis}}
=
\sum_{j\in D}
\log\left[
\sum_{\mathbf z\in\{0,1\}^K}
\widehat\pi_{\mathbf z}^{(0)}
\phi_K(\mathbf W_j;\widehat\Sigma_{\mathbf z}^{(0)})
\right]
\]
where $D$ denotes the set of discrepant SNPs. The goal of penalization is to improve the mixture fit for these SNPs while avoiding substantial deterioration in the overall fit of the model.  For the SNPs that do not affect any trait, their mixture densities are expected to change little under penalization. Hence, we apply VGM-lfdr and retain the set $G_0$ of SNPs identified as affecting at least one trait. Let $L_0$ denote the sum of their log-mixture density.  For each candidate value of $\lambda$, we compute $\mathrm{L_\lambda^{dis}}$ using the same set $D$ and also record $L_\lambda$ using the SNPs in $G_0$. We then quantify the improvement among discrepant SNPs with the corresponding loss in overall mixture fit through
\[
\frac{\mathrm{L}_\lambda^{\text{dis}}-\mathrm{L_0^{dis}}}{L_0-L_\lambda},
\]
and choose the value of $\lambda$ that maximizes this ratio. A simulation study in the Appendix evaluates the performance of lfdr-VGM with this automated choice of $\lambda$ and illustrates how discrepant SNPs can arise in the absence of penalization.


\begin{remark}
In csmGmm, for each trait $k$, the score is modeled using a Gaussian mixture, similar to our approach. However, to capture nonzero effect sizes,
 the method uses pairs of shifted Gaussian components,
$N(-\mu_{km},1)$ and $ N(\mu_{km},1)$,
where $\mu_{km}>0$ for $m=1,\ldots,M$, with $M\geq 1$. Moreover, the mean parameters used to represent nonzero SNP effects may depend on the particular configuration of the vector $Z$. While this construction provides a flexible way to model non-null effects, it also has some practical limitations. In genome-wide applications, significant z-scores can be quite scattered. As a result, even a large z-score may have very small local density under a component such as $N(\mu,1)$, unless the observed value lies sufficiently close to the chosen mean $\mu$. Using multiple mean parameters can help address this issue by covering a wider range of possible effect sizes. However, this introduces another practical question: how should the number of mean parameters be chosen? Increasing the number of components also increases the complexity of the model, which can make the computation challenging when the number of traits is large.
A simple way to reduce this difficulty is to truncate the z-scores and then use only a small number of mean parameters. This is the approach we adopt in both the empirical study and the real data analysis. However, changing the threshold can impact the power and FDR of the method.

Here, we take a more robust approach to modeling the z-scores. We assume the scores follow $N(0,\sigma_k^2)$ across all configurations $Z$ if the corresponding effect size is non-zero. Considering a single parameter $\sigma_k^2$ captures the variability of large z-scores, thereby circumventing issues arising from near-zero density, and also keeps the number of parameters minimal.
\end{remark}



\section{Simulation}\label{sec: simulation}
In this section, we present simulation studies to evaluate the performance of our two proposed sequential tests and compare them with other existing methodologies: PLACO Plus and csmGmm.
PLACO \cite{ray2021pleiotropy} is a statistical method developed to detect ``strict'' pleiotropic SNPs in the setting of two traits. For a given SNP, the PLACO test statistic is defined as the product of the corresponding $z$-scores from the two studies, and its null distribution is derived under the assumption that the two $z$-scores are independent. Later, PLACO Plus \cite{park2025robust} was introduced to accommodate dependence between the $z$-scores across the two traits. However, extending this framework to more than two traits is challenging because of the increasing complexity involved in deriving the asymptotic null distribution of the product statistic.
 We first consider two traits so that the performance of the proposed method can be compared against both PLCO and csmGMM. We further evaluate the performance of the proposed lfdr method against cmGmm for $K$=5 biomarkers.  The general structure of the simulation for any number of traits $K$ is as follows:

\begin{enumerate}
\item The genotype status for each SNP is generated from a trinomial distribution in Hardy-Weinberg equilibrium, with a fixed minor allele frequency of 0.25.

    \item \textbf{Configuration selection for each SNP :} We set the total number of SNPs to $10^5$, with 99\% of SNPs assumed to have no effect on any of the traits. For each of the remaining ``non-null" SNPs, we sample a configuration of $z\in \left\{0,1\right\}^K$ using a specified probability distribution $\pi_{z}$ and correspondingly allow it to be associated with a given subset of the traits. We choose $\pi_z $ proportional to $1/|z|$ so that configurations of a SNP being associated with more traits are less likely than being associated with fewer traits, but it also corresponds to "positive" dependence for SNPs simultaneously being associated with multiple traits. For example, when $K=2$, we selected $\pi_{00}=0.99, \pi_{01}=\pi_{10} = c, \pi_{11}=c/2$ with $c=(1-0.99)2/5$.

    \item \textbf{Trait assignment:} The non-zero effect sizes are generated from $N(0, \sigma_\beta^2)$, where $\sigma_\beta^2 \in \{0.01, 0.05\}$.

    \item \textbf{z-score computation:} For the $i$-th individual and the $j$-th SNP, the $K$-dimensional trait vector is generated from a multivariate normal distribution with mean $S_{ij}\beta_{j\cdot}$, where $S_{ij}$ denotes the SNP value, and a covariance matrix $\Sigma_j$ which corresponds to a residual variance of the biomarkers of 0.1 and a correlation value of 0.3. After generating the data, for each SNP $S_j$, we perform a linear regression of $Y$ on $S_j$ and compute the corresponding $z$-scores.
\end{enumerate}

\begin{figure}[h!]
  \centering
\includegraphics[scale=0.45]{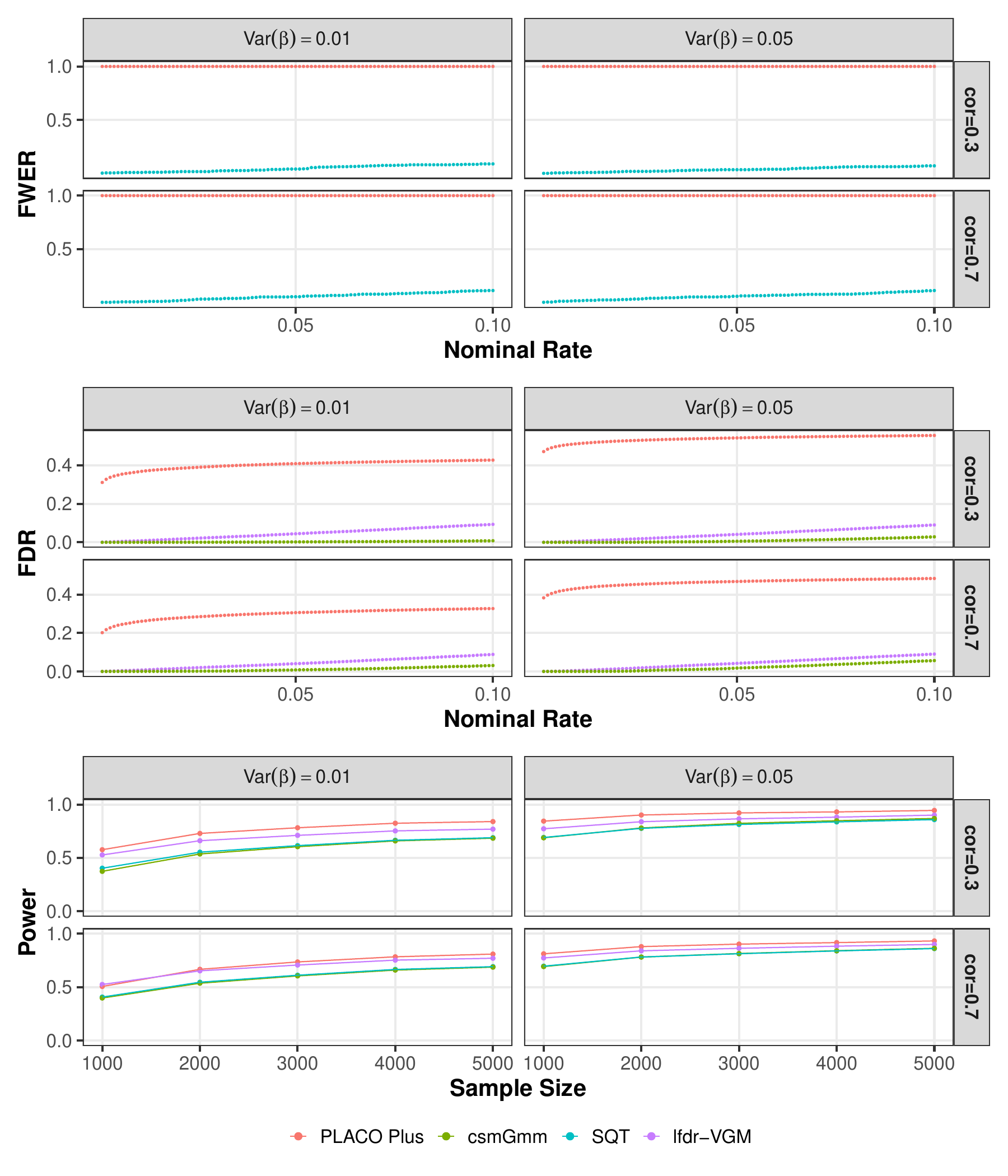}
  \caption{Performance of four testing methods for strict pleiotropy when the number of biomarkers is 2. Non-zero effects for SNPs are simulated from either distribution $N(0, 0.01)$  (left subplots) or $N(0,0.05)$ (right subplots).  Additionally,  for each simulation setting,  estimation errors for effect-sizes across the two traits are allowed to be correlated (with values 0.3 or 0.6) to allow for potential sample overlap across the GWAS studies. The top panel presents FWER of SQR and PLACO Plus, the middle panel presents FDR of lfdr-VGM, csmGmm and PLACO Plus and the bottom panel shows the average proportion of SNPs affecting both traits, identified correctly (power) by the tests across 500 simulated datasets at a nominal level of 0.05. For the type-I error and FDR plots, the sample size for each trait is fixed at 3000.}\label{Fig: PlacoVsSeq}.
\end{figure}

Figure \ref{Fig: PlacoVsSeq} summarizes the performance of alternative methods for $K=2$ when $\sigma_\beta^2=0.01$ and  $\sigma_\beta^2=0.05$.  Overall, both proposed tests, SQT and lfdr-VGM with $\lambda=1$, maintain the corresponding error criteria, i.e., FWER for the former and FDR for the latter, at the desired level. Under the composite null hypothesis, a false positive is considered any SNP that is not associated with both traits but is declared significant. We observe that PLACO, which is designed to maintain a type-I error averaged over the different sub-null hypotheses with weights according to estimates of the proportions of SNPs in different components, can have substantially inflated FWER because it does not guarantee suitable error control under each sub-null hypothesis. In particular, in GWAS, the number of SNPs associated with neither trait is typically much larger than those associated with only one trait; as a result, the average type-I error is dominated by SNPs that are under the global hypothesis. Thus, the FWER for PLACO can be particularly large, as the likelihood of a single false positive result from SNPs that are associated with one trait but not the other is very high. The csmGmm method, by contrast, controls FDR, but it appears more conservative than lfdr-VGM. In terms of power, PLACO was the most powerful of the three methods, which is not surprising given that the method does not provide strict FWER or FDR control. Between lfdr-VGM and csmGmm, lfdr-VGM exhibited higher power across the simulation settings. This is consistent with csmGmm's conservative FDR, which remained substantially below the nominal level and consequently yielded fewer discoveries in our settings.

Next, we considered the setting of $K=5$ biomarkers. From Figure 2, we observe that both methods, SQT and lfdr-VGM, maintain the respective error criteria, i.e., FWER and FDR, under each type of null hypothesis, $H_0^{r}$, $r=1,..5$. We observe that the power of both methods decreases as $r$ increases, and both have the lowest power when $r=5$, i.e., for detecting SNPs associated with all five traits. For the testing of $H_0^{5}$, where we could apply csmGmm, we again observe that the power of the lfdr-VGM method was consistently higher than csmGMM across different scenarios (see Figure 3). We present results from additional simulation studies in the Supplemental Materials to investigate the sensitivity of the lfdr-VGM method to the choice of the tuning parameter ``$\lambda$". We find that across $\lambda=0.01$, $0.1$, and $1$, the method maintains FDR and has comparable power. Additional simulation results for different sample sizes are also presented in Supplemental Materials.

Observe that csmGmm has an FDR that is substantially lower than the nominal level.
We truncate the z-scores, as they are quite scattered, and then model them using a one-mean parameter (both positive and negative numbers of the same magnitude) for each trait.  We observe that, for some SNPs identified as pleiotropic by lfdr-VGM, one of the z-scores is large and the other one is not sufficiently large, such that the smaller pair is not close enough to the estimated mean induced by the alternative and therefore yields almost zero probability under the alternative and consequently is not identified as pleiotropic by csmGmm. Also, csmGmm imposes a constraint on the mean parameters: the mean parameters of the $z$-scores under the configuration where a SNP affects exactly $r$ traits must be larger than those under configurations where the SNP affects fewer traits. Hence, in our setting, if one wants to allow a smaller second mean parameter under the two-trait alternative, this parameter must still be larger than the corresponding mean parameters under the single-trait alternatives. Although this constraint is maintained in csmGmm, it can create difficulties when one component of the $z$-score is large while the other is close to zero. In such cases, the joint density under all configurations $Z\in \{0,1\}\times{0,1}$ may happen to be almost 0, leading to numerical issues.

\begin{figure}
  \centering
\includegraphics[scale=0.4]{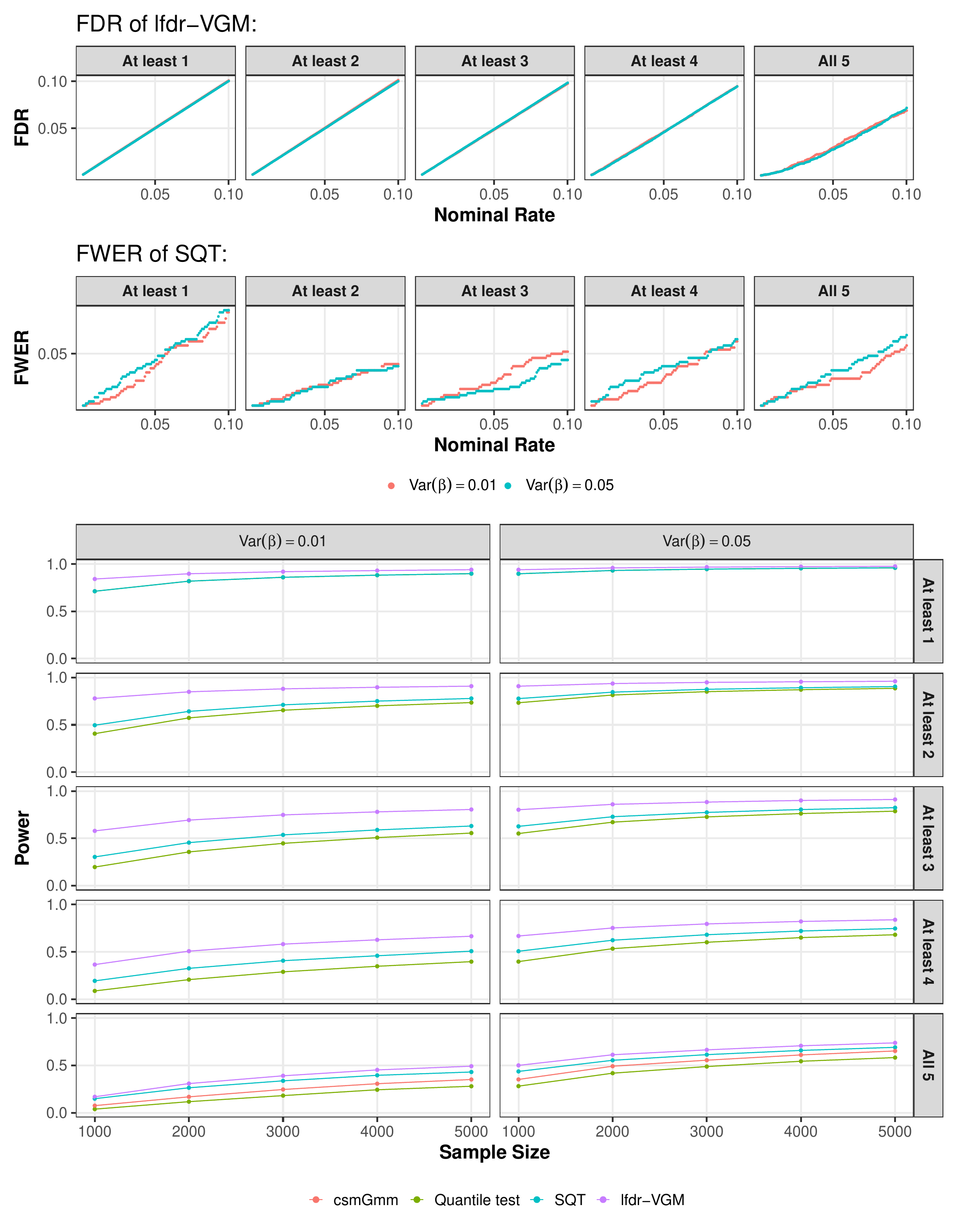}
  \caption{Performance of four testing methods for strict pleiotropy when the number of biomarkers is 5. The top panel shows the empirical FDR of lfdr-VGM, while the middle panel shows the empirical FWER of SQT with a sample size of $3000$. Figures~\ref{Fig:error_plot} in the Appendix reports the corresponding FDR and type I error results for a range of sample sizes. The bottom panel shows the proportion of SNPs identified as affecting at least $r=1,\ldots,5$ traits at the nominal significance level $\alpha=0.05$, averaged over 500 simulated data sets. Nonzero SNP effects are generated from $N(0,0.01)$ (left) and $N(0,0.05)$ (right). The correlation among the random errors is set to $0.03$ to account for potential sample overlap. For lfdr-VGM, the penalty parameter is fixed at $\lambda=1$. The csmGmm method, shown in red, is designed specifically to test whether a SNP affects all traits and therefore appears only in the $r=5$ row.  
  }\label{Fig: Plot_K5}.
\end{figure}

\begin{table*}[t]
\tabcolsep=0pt
\begin{tabular*}{\textwidth}{@{\extracolsep{\fill}}llll|ll|lllll@{\extracolsep{\fill}}}
\hline
RSID       & CHR & POS & Nearest gene                 & A$\beta$ $p$-value & ptau $p$-value & PLACO  & SQT & csmGmm & lfdr-VGM\\
\hline

rs769449  & 19  & 44906745 & APOE                   & 1.69e-321    & 9.59e-95   & 0   &     Identified   & 9.20e-14    & 0   \\
rs117583674 & 7	&			37040693 &ELMO1	& 1.72e-07 &0.007832 &	2.63e-06 &No&0.5831 &0.0334 \\
rs6733839  & 2   & 127135234 &  BIN1               & 1.65e-06     & 1.26e-05   & 9.07e-09  &  No & 0.2417 & 0.005              \\
rs34113929 & 3   & 190902357&  GMNC              & 0.019        & 2.70e-28   & 1.74e-10  &  No & 0.7062 & 0.99          \\
 rs4844610 & 1 & 207629207 & CR1& 1.84e-09
 &0.003016 & 1.10e-07& No & 0.3921 & 0.0008 \\
  rs9430717 & 1 & 9710947	&PIK3CD
&1.12e-07&	0.1503 & 0.000454& No & 0.8581 &0.0266\\
\hline
\end{tabular*}
\caption{Pleiotropic Analysis of AD Biomarkers A$\beta$ and p-tau using multiple alternative methods. The first column gives the RSID of independent SNPs at least 500 kb apart identified by any of the four tests, PLACO Plus, csmGmm, SQT, and lfdr-VGM (with $\lambda=0.01$ as determined by the data-driven method). The next two columns show the chromosome number and the SNP position, and the fourth column gives the nearest gene.  The next three columns provide the $ p$-values of the SNPs from the GWAS summary statistics of A$\beta$ and ptau, as well as the $ p$-values obtained from PLACO, respectively. The last three columns give whether the SNPs are identified by the SQT with FWER at $0.05$ and by csmGmm and lfdr-VGM with FDR at 0.05.}
\label{tab:PLACO}
\end{table*}

\begin{table}[ht]
\centering
\begin{tabular}{cccccc}
\hline
Probability & Abeta & ptau & NFL & Snap25 & Neurogranin \\
\hline
$9.974701\times 10^{-1}$ & 0 & 0 & 0 & 0 & 0 \\
$1.315137\times 10^{-3}$ & 0 & 0 & 1 & 0 & 0 \\
$4.067888\times 10^{-4}$ & 0 & 0 & 0 & 1 & 1 \\
$3.959063\times 10^{-4}$ & 0 & 0 & 0 & 0 & 1 \\
$2.495239\times 10^{-4}$ & 0 & 0 & 0 & 1 & 0 \\
$4.856971\times 10^{-5}$ & 1 & 1 & 1 & 1 & 0 \\
$4.310618\times 10^{-5}$ & 0 & 1 & 0 & 0 & 0 \\
$2.558073\times 10^{-5}$ & 0 & 1 & 1 & 0 & 0 \\
$1.987200\times 10^{-5}$ & 1 & 1 & 0 & 1 & 0 \\
$3.535469\times 10^{-6}$ & 1 & 0 & 1 & 1 & 0 \\
$3.463866\times 10^{-6}$ & 0 & 1 & 1 & 1 & 0 \\
$3.010157\times 10^{-6}$ & 1 & 1 & 1 & 0 & 0 \\
$2.977961\times 10^{-6}$ & 1 & 0 & 0 & 1 & 0 \\
$2.668988\times 10^{-6}$ & 0 & 1 & 0 & 1 & 0 \\
$2.572893\times 10^{-6}$ & 0 & 0 & 1 & 1 & 0 \\
$2.538858\times 10^{-6}$ & 1 & 1 & 0 & 0 & 0 \\
$1.206378\times 10^{-6}$ & 1 & 0 & 0 & 0 & 0 \\
$1.090052\times 10^{-6}$ & 0 & 0 & 1 & 0 & 1 \\
$1.053647\times 10^{-6}$ & 1 & 0 & 1 & 0 & 0 \\
$3.903386\times 10^{-7}$ & 0 & 1 & 0 & 0 & 1 \\
$2.217845\times 10^{-7}$ & 0 & 0 & 1 & 1 & 1 \\
$2.027155\times 10^{-7}$ & 1 & 0 & 0 & 0 & 1 \\
$1.190120\times 10^{-7}$ & 0 & 1 & 0 & 1 & 1 \\
$1.051288\times 10^{-7}$ & 0 & 1 & 1 & 0 & 1 \\
$8.215690\times 10^{-8}$ & 1 & 0 & 0 & 1 & 1 \\
$6.688000\times 10^{-8}$ & 1 & 0 & 1 & 0 & 1 \\
$6.661116\times 10^{-8}$ & 1 & 1 & 0 & 0 & 1 \\
$2.713584\times 10^{-8}$ & 0 & 1 & 1 & 1 & 1 \\
$2.042110\times 10^{-8}$ & 1 & 0 & 1 & 1 & 1 \\
$1.988203\times 10^{-8}$ & 1 & 1 & 0 & 1 & 1 \\
$1.748655\times 10^{-8}$ & 1 & 1 & 1 & 0 & 1 \\
$4.380559\times 10^{-9}$ & 1 & 1 & 1 & 1 & 1 \\
\hline
\end{tabular}
\caption{Estimated probabilities of latent configurations sorted in decreasing order with penalty=10. }
\label{tab:estimated_pi_sorted}
\end{table}

\begin{table}[htbp]
\centering

\scriptsize
\setlength{\tabcolsep}{4pt}
\renewcommand{\arraystretch}{1.1}

\resizebox{\textwidth}{!}{
\begin{tabular}{lccccc}
\toprule
 & Amyloid-$\beta$ & p-tau & NfL & SNAP-25 & Neurogranin \\
\midrule
Amyloid-$\beta$
& \cellcolor{gray!20} 1.0000
& \cellcolor{blue!15} -0.2500
& \cellcolor{blue!15}  0.0004
& \cellcolor{blue!15}  0.0004
& \cellcolor{blue!15}  0.0013
\\

p-tau
& \cellcolor{red!15} -0.9331
& \cellcolor{gray!20} 1.0000
& \cellcolor{blue!15}  0.0015
& \cellcolor{blue!15} -0.0011
& \cellcolor{blue!15} -0.0007
\\

NfL
& \cellcolor{red!15} -0.1030
& \cellcolor{red!15}  0.0034
& \cellcolor{gray!20} 1.0000
& \cellcolor{blue!15} -0.0772
& \cellcolor{blue!15} -0.0130
\\

SNAP-25
& \cellcolor{red!15}  0.1071
& \cellcolor{red!15} -0.0716
& \cellcolor{red!15} -0.3803
& \cellcolor{gray!20} 1.0000
& \cellcolor{blue!15}  0.6193
\\

Neurogranin
& \cellcolor{red!15}  0.0588
& \cellcolor{red!15}  0.0768
& \cellcolor{red!15}  0.2585
& \cellcolor{red!15}  0.5971
& \cellcolor{gray!20} 1.0000
\\
\bottomrule
\end{tabular}
}
\caption{The correlation structure among the z-scores of five traits. Upper triangle: $\rho^e$ (blue). Lower triangle: $\rho^b$ (red).}
\label{tab:cor_combined}
\end{table}

\begin{table}[!htbp]
\centering
\scriptsize
\setlength{\tabcolsep}{3pt}
\renewcommand{\arraystretch}{0.8}
\resizebox{\textwidth}{!}{%
\begin{tabular}{lrrlccccccc}
\hline
rsid & chr & pos & Nearest gene & A$\beta$ & p-tau & NFL & SNAP-25 & Neurogranin & \# traits  lfdr-VGM & \# traits SQT \\
\hline
rs429358 & 19 & 44908684 & APOE & 3.39e-38 &9.75e-16& 0.00323& 3.13e-176 0&.433& 4 & 4 \\
rs2972558 & 19 &44852884&	NECTIN2 & 9.28e-17	&	0.01005 & 0.000638& 6.13e-41
&0.574 & 4 & 3\\
rs73045691 & 19 & 44937272 & APOC4 & 5.60e-07	&	0.001608& 0.000337& 9.91e-20 & 0.134	& 4 & 3 (LD with rs77301115)\\
rs8103315 & 19 & 44750911 & BCL3 & 7.29e-14 & 2.07e-05 & 3.67e-02 & 1.92e-30 & 7.72e-01 & 3 & 3\\
rs157585	& 19 & 44894255&TOMM40 & 1.35e-44	&1.10e-14 & 0.405 & 2.07e-46 & 0.956 & 3 & 3\\
rs28795074	& 19 &44933400 &APOC1P1 & 6.54e-07	& 	0.004795 &0.153 & 1.37e-06& 0.507 & 3 & 0\\
rs62117205 & 19 & 44752009&	BCL3 & 9.48e-09	& 3.391	6.975e-4&0.964 & 1.29e-10 & 0.922 & 3 & 2\\

\hline
rs2070116 & 22 & 42210251 & OGFRP1 & 1.13e-01 & 9.01e-01 & 4.13e-02 & 4.51e-09 & 1.57e-02 & 2 & 1 \\
rs4803759 & 19 & 44824202 & BCAM & 2.51e-11 & 1.02e-01 & 3.75e-01 & 8.20e-25 & 3.02e-01 & 2 &2 \\
rs4803824 & 19 & 45448020 & ERCC1 & 9.18e-01 & 5.13e-01 & 6.92e-01 & 2.41e-08 & 4.66e-04 & 2 & 0 \\
rs1671152 & 19 & 55014977 & GP6-AS1 & 3.74e-01 & 7.51e-01 & 7.01e-01 & 4.55e-07 & 1.09e-13 & 2 & 2\\
rs7210399 & 17 & 29453198 & TAOK1 & 1.76e-01 & 5.96e-02 & 8.06e-01 & 1.96e-06 & 8.24e-09 & 2 & 2 (LD with rs1967556)\\
rs1298104 & 16 & 4386444 & TFAP4 & 4.00e-01 & 6.42e-02 & 9.06e-01 & 1.34e-08 & 1.21e-02 & 2 & 0\\
rs6580981 & 12 & 54329244 & COPZ1 & 7.43e-01 & 4.93e-01 & 6.54e-01 & 2.71e-03 & 1.25e-07 & 2 & 0\\
\hline
rs10822145 & 10 & 63174788 & JMJD1C & 9.95e-01 & 1.72e-01 & 1.19e-01 & 5.47e-17 & 3.85e-13 & 2 & 2 (LD with rs7896518)\\
rs113422568 & 10 & 102578181 & SUFU & 4.77e-01 & 3.49e-01 & 6.59e-02 & 1.82e-04 & 5.54e-06 & 2 & 0\\
rs61469632 & 9 & 132986603 & GFI1B & 8.95e-01 & 9.53e-02 & 2.28e-01 & 1.69e-11 & 1.74e-03 & 2 & 1\\
rs467369 & 9 & 134040643 & BRD3 & 7.28e-01 & 5.53e-01 & 7.06e-01 & 4.39e-07 & 6.68e-03 & 2 & 0 \\
rs4734879 & 8 & 105570896 & ZFPM2 & 1.89e-02 & 6.66e-01 & 7.77e-01 & 4.42e-09 & 6.14e-13 & 2 & 2\\
\hline
rs193117411 & 7 & 106685425 & CCDC71L & 4.73e-01 & 5.96e-01 & 6.50e-01 & 1.14e-08 & 2.78e-02 & 2 & 0\\
rs342296 & 7 & 106732457 & LINC02577 & 6.20e-03 & 5.46e-01 & 1.53e-01 & 2.57e-172 & 1.79e-12 & 2 & 2\\
rs9389269 & 6 & 135106021 & HBS1L & 6.39e-01 & 7.78e-02 & 3.67e-01 & 2.28e-05 & 3.68e-09 & 2 & 1\\
rs115318824 & 5 & 89001177 & MEF2C-AS1 & 1.41e-01 & 4.07e-01 & 4.49e-01 & 1.49e-05 & 5.58e-03 & 2 & 0\\
rs10900809 & 5 & 132490630 & IRF1 & 1.27e-01 & 6.36e-01 & 5.88e-01 & 5.67e-06 & 4.11e-10 & 2 & 2\\
\hline
rs1354034 & 3 & 56815721 & ARHGEF3 & 5.25e-01 & 3.53e-01 & 1.93e-01 & 2.23e-06 & 5.24e-14 & 2 & 2 \\
rs1501600 & 3 & 190903230 & GMNC & 2.32e-02 & 8.43e-29 & 2.91e-03 & 9.53e-03 & 1.01e-01 & 2 & 1\\
rs6733839 & 2 & 127135234 & BIN1 & 1.65e-06 & 1.26e-05 & 2.61e-01 & 6.01e-01 & 9.22e-01 & 2 & 0\\
rs750439 & 1 & 156460957 & MIR9-1HG & 7.49e-01 & 7.36e-01 & 4.08e-01 & 5.55e-11 & 5.67e-03 & 2 & 1\\
rs4844610 & 1 & 207629207 & CR1 & 1.84e-09 & 3.02e-03 & 3.20e-01 & 6.37e-01 & 7.99e-01 & 2 & 1\\
\hline
\end{tabular}%
}
\caption{SNPs affecting at least two traits related to Alzheimers' identified by the lfdr-VGM and SQT are presented. The first column lists the SNP's RSID. The following three columns provide the chromosome number, position of the SNP, and the nearest gene. From the fifth to the ninth columns, the $p$-values of five traits corresponding to each SNP are presented. The next column gives the minimum number of traits the SNP in the first column is associated with, as examined by lfdr-VGM ($\lambda=10$ chosen by the data-based method). The next column presents the minimum number of traits the SNP in the first row is associated with, as identified by the SQT. If the SQT identifies a different SNP that is in LD with the SNP in the first column, the identified SNP is written in open brackets. Note that the SNPs are obtained after clumping and thresholding with a distance parameter of 500 kb and an LD threshold of 0.01.}
\label{tab:bio_seq_lfdr_results}
\end{table}

\section{Data Example}

We apply SQT and lfdr-VGM to GWAS summary statistics for biomarkers related to Alzheimer's disease (AD), with PLACO Plus and csmGmm included for comparison when applicable. Cerebrospinal fluid (CSF) amyloid-$\beta$ (A$\beta$42) and phosphorylated tau (p-tau) reflect key pathological changes in AD: A$\beta$42 decreases in CSF as amyloid aggregates in the brain, whereas p-tau increases with neuronal injury (see \cite{jack2013biomarker,jack2013update}). GWAS summary statistics for A$\beta$42 and p-tau were obtained from the European Alzheimer \& Dementia Biobank, comprising 31 cohorts and 13,116 individuals, including 8,074 in the discovery sample \cite{jansen2022genome}.

We first consider the two-biomarker analysis and apply PLACO Plus, SQT, lfdr-VGM, and csmGmm to identify variants associated with both A$\beta$42 and p-tau. PLACO Plus is applied at the genome-wide threshold $5\times10^{-8}$, SQT controls genome-wide FWER at 0.05, and lfdr-VGM and csmGmm are applied at FDR 0.05. The data-driven procedure selects $\lambda=0.01$ for lfdr-VGM, yielding estimated probabilities 0.999, $7.33\times10^{-7}$, $7.16\times10^{-5}$, and $6.54\times10^{-5}$ for configurations $(0,0)$, $(1,0)$, $(0,1)$, and $(1,1)$, respectively. The estimated correlations are $\widehat\rho_e=-0.25$ and $\widehat\rho_b=-0.933$, indicating correlation arising from sample overlap and strong correlation between nonzero effects across the two biomarkers. For all methods, we apply clumping and thresholding using $r^2=0.01$ and a 500 kb window to retain independent lead SNPs.

Table~\ref{tab:PLACO} reports independent lead SNPs identified by at least one method. All four methods identify the chromosome 19 APOE region as showing pleiotropic association with both biomarkers. PLACO Plus and lfdr-VGM also identify additional loci, although their findings differ. Both identify a chromosome 2 locus near BIN1. PLACO Plus additionally identifies a chromosome 3 locus for which the index SNP is strongly associated with p-tau ($p=2.7\times10^{-28}$) but only modestly associated with A$\beta$ ($p=0.019$). In contrast, lfdr-VGM identifies one locus on chromosome 7 and two on chromosome 1 that are not genome-wide significant under PLACO Plus. The estimated probability of the SNP affecting A$\beta$ is substantially smaller than the probability of a SNP affecting both traits, leading lfdr-VGM to identify rs9430717.

We next extend the analysis to five AD-related biomarkers by including neurogranin, neurofilament light (NFL), and SNAP-25. CSF neurogranin is elevated in AD-related dementia and MCI due to AD \cite{kvartsberg2015cerebrospinal,wellington2016increased,lista2017synaptic}; NFL, a marker of axonal injury, is increased in CSF and plasma \cite{giacomucci2022plasma,moscoso2021longitudinal,weston2017serum}; and CSF SNAP-25 fragments are elevated even at early stages of AD \cite{brinkmalm2014snap,zhang2018cerebrospinal}. NFL GWAS summary statistics were obtained from 11 European cohorts in the Rotterdam and Rhineland Studies, comprising 18,532 individuals \cite{ahmad2024genome}. SNAP-25 and neurogranin GWAS data were obtained from the UK Biobank, with the SNAP-25 discovery sample including 47,745 individuals of European ancestry \cite{loya2025scalable}.

We apply SQT and lfdr-VGM to identify SNPs and the minimum number of biomarkers with which they are associated. The data-driven procedure selects $\lambda=10$ for lfdr-VGM. Table~\ref{tab:estimated_pi_sorted} reports the estimated latent configuration probabilities. As expected, the null configuration has the largest probability, approximately 0.997. Notably, the configuration corresponding to joint effects on A$\beta$, p-tau, NFL, and SNAP-25 has a relatively high estimated probability, exceeding those of a subset of at least two traits. The SNAP-25--neurogranin configuration also has a comparatively high probability, whereas configurations in which neurogranin is jointly affected with other biomarkers are much less common. These patterns indicate stronger estimated shared genetic association for A$\beta$, p-tau, NFL, and SNAP-25, and between SNAP-25 and neurogranin.

Table~\ref{tab:cor_combined} reports the estimated correlations between random errors in the upper triangle and between nonzero effect sizes in the lower triangle. As expected from overlapping GWAS samples, the random errors are correlated for A$\beta$ and p-tau and for SNAP-25 and neurogranin. The estimated nonzero effects are also strongly correlated between A$\beta$ and p-tau and between neurogranin and SNAP-25.

\begin{figure}[h!]
  \centering
  \includegraphics[scale=0.32]{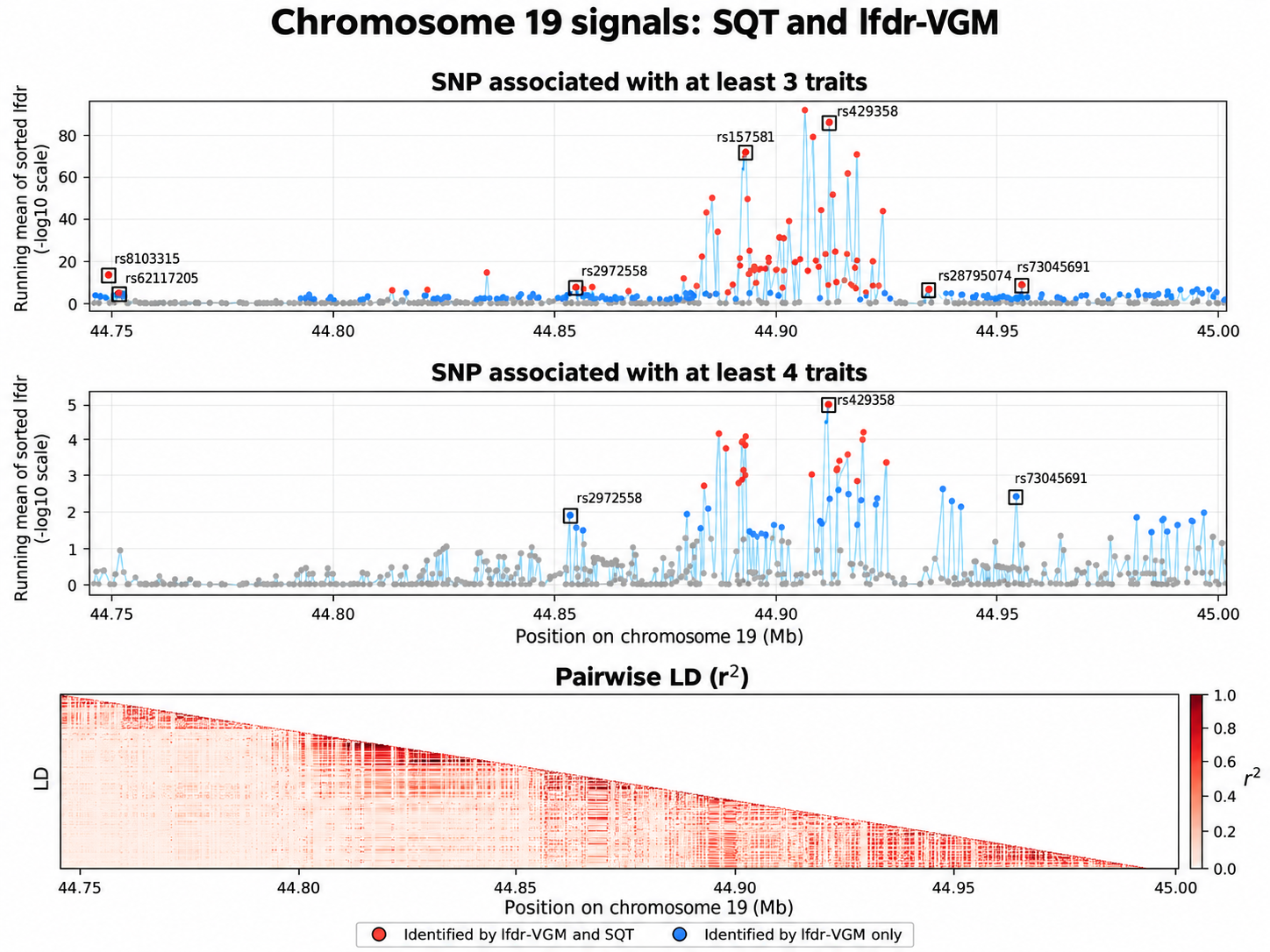}
  \caption{SNPs identified by either SQT or lfdr-VGM across chromosome 19.
  The top panel shows SNPs identified as being associated with at least three traits, with red points indicating SNPs identified by both methods and blue points indicating SNPs identified by lfdr-VGM only.
  The middle panel similarly shows SNPs identified as being associated with at least four traits.
  In both panels, the running mean of the ordered lfdr values, shown on the $-\log_{10}$ scale, is displayed as a blue curve for the corresponding hypothesis.
  The bottom panel displays linkage disequilibrium (LD), measured by pairwise $r^2$, across the genomic positions.}
  \label{Fig: dataplot}
\end{figure}

Figure~\ref{Fig: dataplot} compares SQT and lfdr-VGM in the APOE region. lfdr-VGM identifies more SNPs as affecting at least three and at least four traits. The additional discoveries at the four-trait level are partly explained by the relatively large estimated probability of configuration $(1,1,1,1,0)$, corresponding to joint effects on A$\beta$, p-tau, NFL, and SNAP-25, compared with its size-three subsets. Table~\ref{tab:bio_seq_lfdr_results} reports the independent SNPs identified after clumping and thresholding. The APOE-region findings are consistent with previous reports linking rs7412 and rs429358 to Alzheimer's disease (see \cite{kulminski2020genetic,ware2020considering,guo2019genome}). Both SQT and lfdr-VGM identify rs429358 as affecting at least four traits and rs7412 as affecting at least three. lfdr-VGM additionally identifies rs2972558, which is in LD with rs7412, as affecting at least four traits. Overall, SQT identifies SNPs on chromosomes 3, 5, 7, 8, 10, 17, and 19 as affecting at least two traits, whereas lfdr-VGM identifies independent SNPs across chromosomes 1, 2, 3, 5, 6, 7, 8, 9, 10, 12, 16, 17, 19, and 22.

\section{Discussion}
In this article, we proposed two methods, SQT and lfdr-VGM, for identifying pleiotropic SNPs using Wald test statistics. A key novelty of both methods is that, beyond detecting pleiotropy, they can determine the minimum number of traits a SNP is associated with at a given significance level.
SQT has several practical advantages. It is simple and computationally efficient to implement, accommodates both independent and correlated traits, can be applied to an arbitrary number of traits, unlike methods such as PLACO that are restricted to a smaller number of traits, and requires only summary-level test statistics rather than individual-level data.
The lfdr-VGM approach involves estimation of the underlying mixture model and is therefore computationally more involved than SQT. However, it provides a substantial power gain. While SQT controls FWER, lfdr-VGM controls FDR, allowing it to identify more pleiotropic variants while maintaining appropriate error control. In addition, lfdr-VGM explicitly models both the correlation among random errors across traits, which can arise from sample overlap, and the correlation among nonzero genetic effects. Incorporating these dependence structures into the joint distribution of the Wald statistics further improves its ability to detect pleiotropic signals.

The performance of csmGmm depends on the number of nonzero mean parameters used and has power comparable to SQT in our current setting. It is also conservative in controlling the FDR at the nominal level (one mean parameter is used for each trait in the latent state where a SNP affects all traits), and is therefore substantially less powerful than the lfdr-VGM method.

Another important distinction between csmGmm and lfdr-VGM concerns model complexity. For csmGmm, we consider one set of mean parameters for each trait under each latent configuration. Consequently, the number of mean parameters grows as $ O(K^2) $. In contrast, the lfdr-VGM method uses a single variance parameter for each trait. Thus, as the number of traits increases, lfdr-VGM requires substantially fewer parameters and can be computationally less intensive than csmGmm.

Even though lfdr-VGM maintains FDR, its performance can be sensitive to the estimated probabilities in situations where a configuration involving a larger number of traits is assigned a substantially higher probability than a lower-order configuration corresponding to a subset of those traits, and the lower-order configuration is not truly absent but is represented by only a small number of SNPs. We attempted to mitigate this issue by introducing a penalty in the estimation of the configuration probabilities. Another limitation is that the method can still become computationally challenging as the number of traits increases. An important direction for future work is therefore to develop a simplified and more scalable version of lfdr-VGM for applications involving a large number of traits.

\bibliographystyle{apalike}
\bibliography{Pleiotropy_ref}


\appendix
\section{Appendix}
\subsection{Additional Simulation Results}

In this section, we present additional figures depicting the power of lfdr-VGM and SQT and compare these methods with other existing methods. The simulation procedure is the same as prescribed in Section \ref{sec: simulation}.

Recall that to perform lfdr-VGM, one intermediate step is to estimate the probabilities of the latent variable $Z \in \{0,1\}^K$, with $K$ being the number of traits. To estimate these probabilities, we introduced penalized likelihood. Figure \ref{Fig: Plot_K5} in Section \ref{sec: simulation} presents the power of lfdr-VGM with penalty $\lambda=1$. That figure also presents the FDR of lfdr-VGM and the FWER of SQT for a sample size of 3000. Figure \ref{Fig:error_plot} presents the FDR of lfdr-VGM and the FWER of SQT for sample sizes 1000 to 5000.

\begin{figure}
  \centering
\includegraphics[scale=0.37]{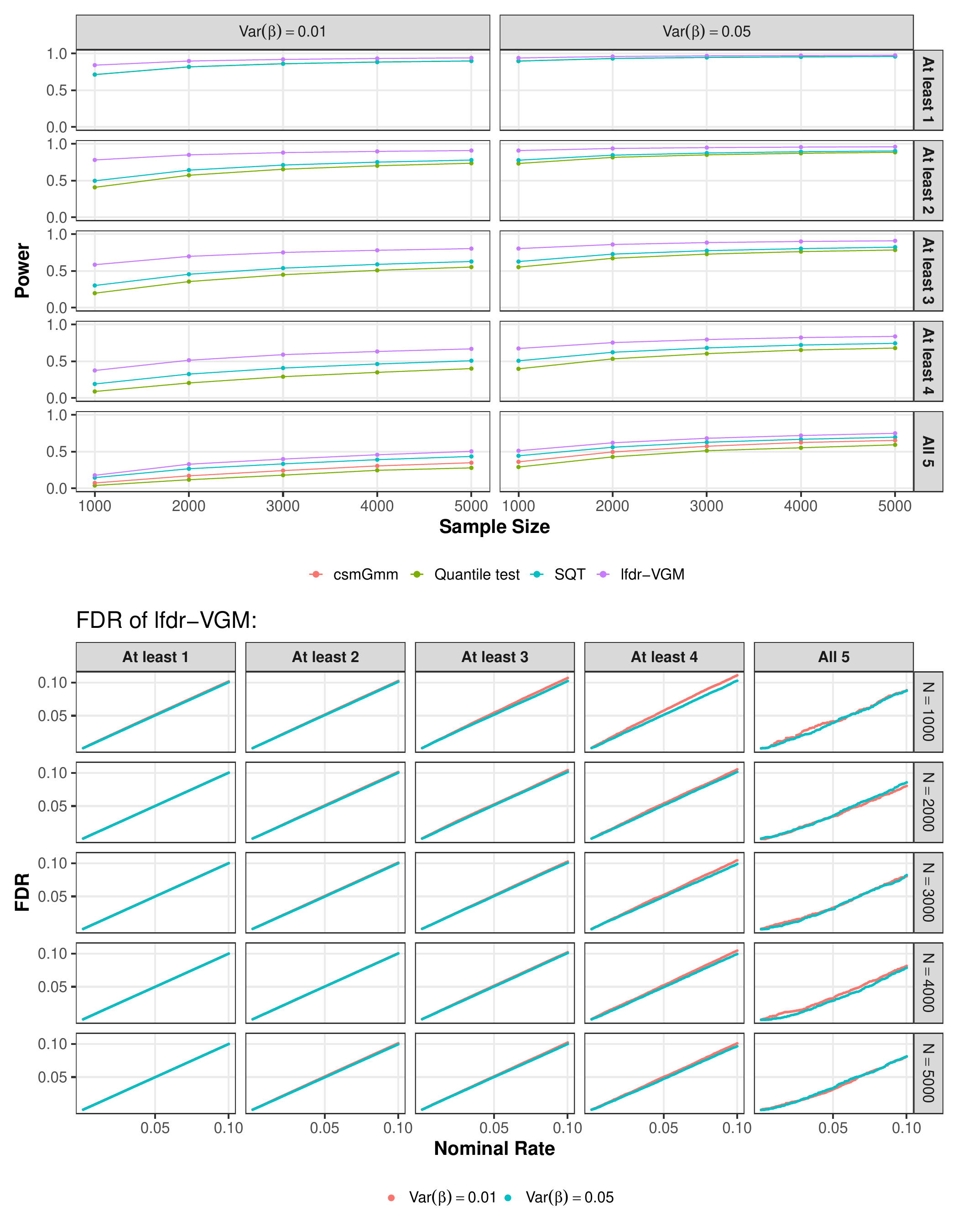}
  \caption{Performance of four methods to identify SNPs affecting multiple traits. The top panel shows the empirical FDR of lfdr-VGM with penalty parameter $\lambda=0.01$ for different sample sizes presented in different rows. The two colors denote the effect-size variance used in the simulation.  The bottom panel shows the proportion of SNPs identified as affecting at least $r=1,\ldots,5$ traits at the nominal significance level $\alpha=0.05$, averaged over 500 simulated data sets, for four testing methods. The csmGmm method, shown in red, is designed specifically to test whether a SNP affects all traits and therefore appears only in the $r=5$ row.
  }\label{Fig: Plot_K5_pen_0p01}.
\end{figure}

\begin{figure}
  \centering
\includegraphics[scale=0.37]{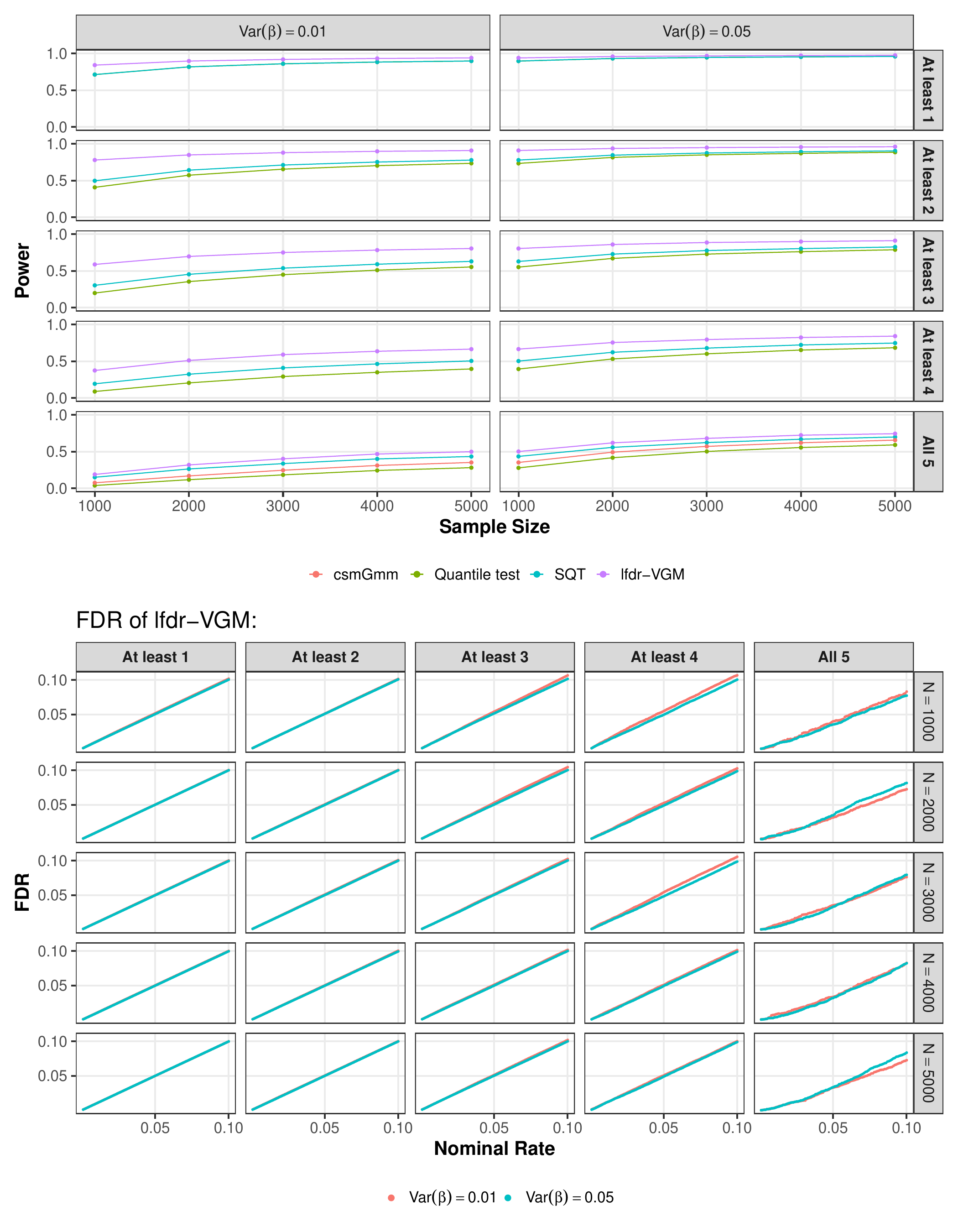}
  \caption{Performance of four methods to identify SNPs affecting multiple traits. The top panel shows the empirical FDR of lfdr-VGM with penalty parameter $\lambda=0.1$ for different sample sizes presented in different rows. The two colors denote the effect-size variance used in the simulation.  The bottom panel shows the proportion of SNPs identified as affecting at least $r=1,\ldots,5$ traits at the nominal significance level $\alpha=0.05$, averaged over 500 simulated data sets, for four testing methods. The csmGmm method, shown in red, is designed specifically to test whether a SNP affects all traits and therefore appears only in the $r=5$ row.
  }\label{Fig: Plot_K5_pen_0p1}.
\end{figure}

\begin{figure}[h]
  \centering
\includegraphics[scale=0.45]{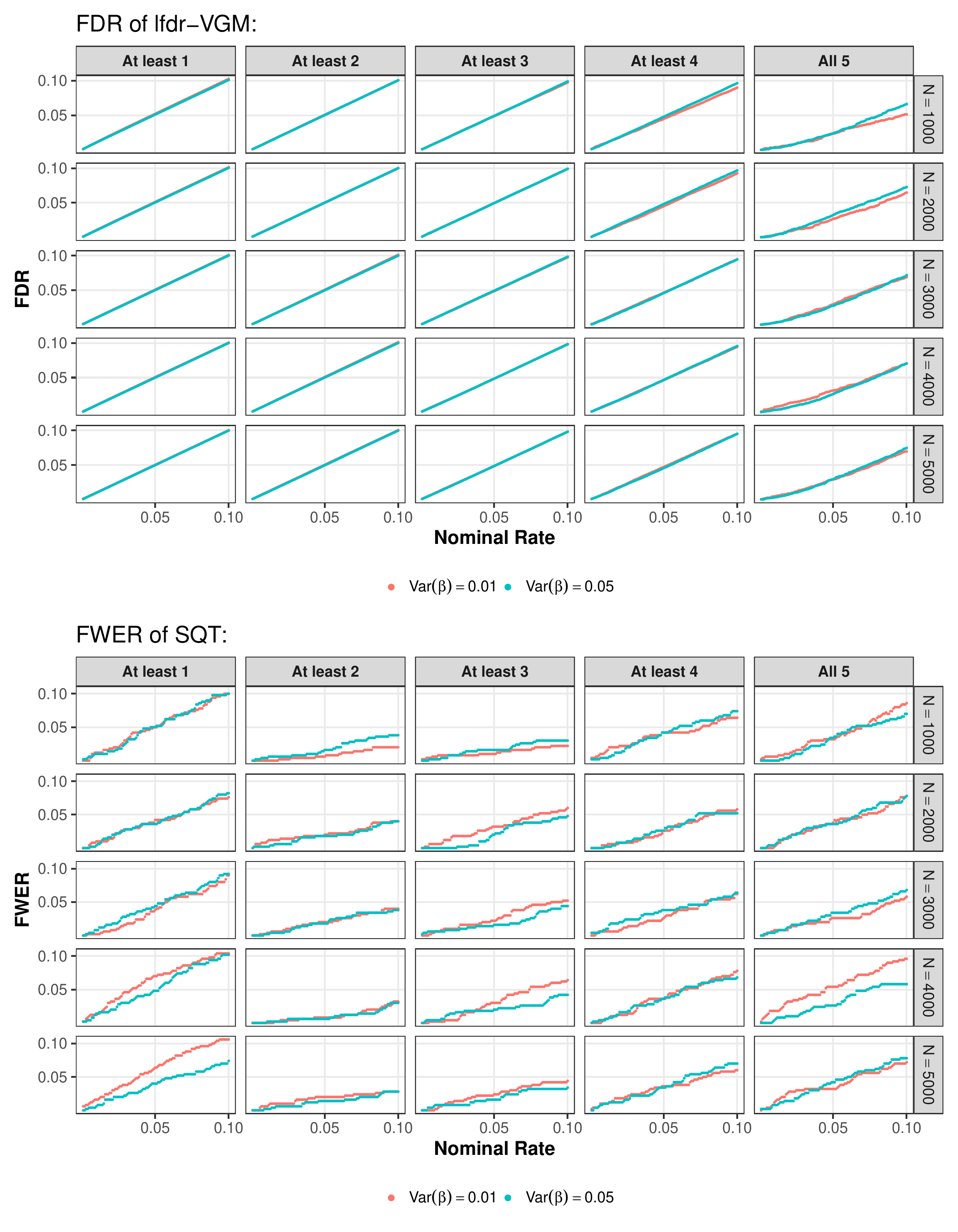}
\caption{Error control by two methods in identifying SNPs affecting multiple traits. The top panel shows FDR using lfdr-VGM with $\lambda=1$, while the bottom one shows empirical FWER of SQT for testing whether a SNP affects at least $r=1,...,5$ traits.  The x-axis presents the nominal values from 0 to 0.1. Five sample sizes are used to obtain the z-scores presented along different rows, while different colors show the different values of variances of non-zero effect sizes.}
\label{Fig:error_plot}
\end{figure}

Next, we compare the power and FDR of lfdr-VGM if we reduce the penalty.
We present Figures \ref{Fig: Plot_K5_pen_0p01} and \ref{Fig: Plot_K5_pen_0p1}, which illustrate the power and FDR of lfdr-VGM with penalty $\lambda=0.01$ and 0.1, respectively, for sample sizes 1000, 2000, 3000, 4000 and 5000 and variance on non-zero effect sizes 0.01, 0.05 and 0.1. We observe that lfdr-VGM has higher power than other methods in these two figures. Also, it maintains FDR at a nominal level even after reducing $\lambda$ to 0.1 and 0.01. However, when we increase $ \lambda$ to 1, the FDR for some null hypotheses is smaller than under a smaller penalty.

\begin{figure}[h!]
  \centering
\includegraphics[scale=0.4]{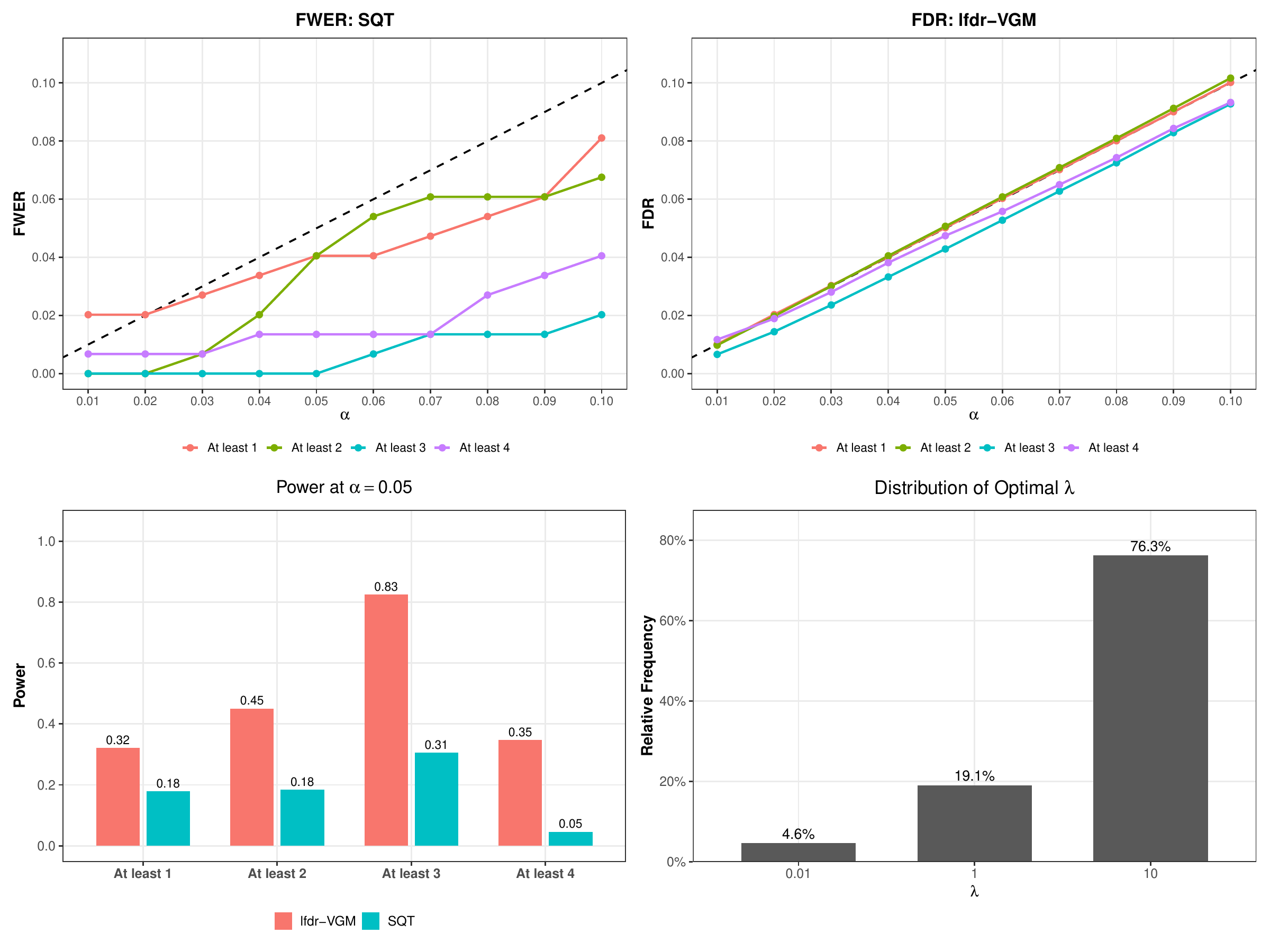}
  \caption{Performance of SQT and lfdr-VGM, simulated using parameter estimates from the data. The top panel shows FWER (left) and FDR (right) maintained by SQT and lfdr-VGM, respectively, under four hypotheses. The bottom-left panel shows the power of SQT and lfdr-VGM for optimal $\lambda$, while the right panel shows the frequency of optimal $\lambda$ based on 150 simulated datasets with $7\times 10^6$ SNPs (similar to the data).}\label{Fig: plot_using_data_estimates}.
\end{figure}

We also investigate the performance of lfdr-VGM with $\lambda$ chosen by a data-driven method, as depicted in Section \ref{lfdr}. We generate z-scores using the estimated parameters obtained by modeling the five AD-related traits with $\lambda=0$.  Figure \ref{Fig: plot_using_data_estimates} presents the performance of automatic lfdr-VGM and SQT.

\subsection{Data Analysis without penalty}
Table \ref{tab:seq_lfdr_independent_snps_pen_0} presents independent SNPs with an LD threshold of 0.01 and at least 500 kb apart, identified by lfdr-VGM without penalty. We notice four SNPs that are associated with at least three traits even though the p-values from the GWAS summary are not sufficiently low for at least 3. This is because the estimated probability that a SNP affects A$\beta$, p-tau, and SNAP25 is significantly higher than the probability that it affects a subset of two, as shown in Table \ref{tab:estimated_pi_sorted_pen0}.
\begin{table}[htbp]
\centering
\begin{tabular}{cccccc}
\hline
A$\beta$ & pTau & NFL & SNAP25 & Neurogranin & Estimated Probability \\
\hline
0 & 0 & 0 & 0 & 0 & $9.97474\times 10^{-1}$ \\
0 & 0 & 1 & 0 & 0 & $1.31415\times 10^{-3}$ \\
0 & 0 & 0 & 1 & 1 & $4.08962\times 10^{-4}$ \\
0 & 0 & 0 & 0 & 1 & $3.96246\times 10^{-4}$ \\
0 & 0 & 0 & 1 & 0 & $2.49383\times 10^{-4}$ \\
1 & 1 & 1 & 1 & 0 & $5.74174\times 10^{-5}$ \\
0 & 1 & 0 & 0 & 0 & $4.36340\times 10^{-5}$ \\
1 & 1 & 0 & 1 & 0 & $2.86119\times 10^{-5}$ \\
0 & 1 & 1 & 0 & 0 & $2.73186\times 10^{-5}$ \\
1 & 0 & 1 & 1 & 0 & $4.91866\times 10^{-7}$ \\
0 & 0 & 1 & 0 & 1 & $7.70435\times 10^{-8}$ \\
1 & 0 & 0 & 1 & 0 & $5.08188\times 10^{-8}$ \\
0 & 0 & 1 & 1 & 0 & $3.21084\times 10^{-9}$ \\
0 & 0 & 1 & 1 & 1 & $5.27477\times 10^{-10}$ \\
1 & 1 & 0 & 0 & 0 & $6.89983\times 10^{-14}$ \\
0 & 1 & 0 & 0 & 1 & $1.00000\times 10^{-14}$ \\
0 & 1 & 0 & 1 & 0 & $1.00000\times 10^{-14}$ \\
0 & 1 & 0 & 1 & 1 & $1.00000\times 10^{-14}$ \\
0 & 1 & 1 & 0 & 1 & $1.00000\times 10^{-14}$ \\
0 & 1 & 1 & 1 & 0 & $1.00000\times 10^{-14}$ \\
0 & 1 & 1 & 1 & 1 & $1.00000\times 10^{-14}$ \\
1 & 0 & 0 & 0 & 0 & $1.00000\times 10^{-14}$ \\
1 & 0 & 0 & 0 & 1 & $1.00000\times 10^{-14}$ \\
1 & 0 & 0 & 1 & 1 & $1.00000\times 10^{-14}$ \\
1 & 0 & 1 & 0 & 0 & $1.00000\times 10^{-14}$ \\
1 & 0 & 1 & 0 & 1 & $1.00000\times 10^{-14}$ \\
1 & 0 & 1 & 1 & 1 & $1.00000\times 10^{-14}$ \\
1 & 1 & 0 & 0 & 1 & $1.00000\times 10^{-14}$ \\
1 & 1 & 0 & 1 & 1 & $1.00000\times 10^{-14}$ \\
1 & 1 & 1 & 0 & 0 & $1.00000\times 10^{-14}$ \\
1 & 1 & 1 & 0 & 1 & $1.00000\times 10^{-14}$ \\
1 & 1 & 1 & 1 & 1 & $1.00000\times 10^{-14}$ \\
\hline
\end{tabular}
\caption{Estimated probabilities for biomarker association configurations without penalty.}
\label{tab:estimated_pi_sorted_pen0}
\end{table}

\begin{table}[!htbp]
\centering
\scriptsize
\setlength{\tabcolsep}{3pt}
\renewcommand{\arraystretch}{0.8}
\resizebox{\textwidth}{!}{%
\begin{tabular}{lrrlccccccc}
\hline
RSID & CHR & POS & Nearest gene
& A$\beta$ $p$-value
& p-tau $p$-value
& NFL $p$-value
& SNAP-25 $p$-value
& Neurogranin $p$-value
& SQT
& lfdr-VGM \\
\hline

rs429358 & 19 & 44908684 & APOE & 3.39e-38 &9.75e-16& 0.00323&  0& 0.433& 4 & 4 \\

rs2107964 & 7 &2823940& GNA12 & 0.0059 & 3.29e-05& 0.046 & 0.00252 & 0.046 & 0& 3\\
rs9260061 &6 & 29940714 & LOC105375010 &0.003522 & 0.001158& 0.886& 1.73e-06& 0.13 &  2 with rs342296 (LD) & 3\\
rs6733839 & 2 & 127135234&BIN1 & 1.65e-06&1.26e-05&0.261 & 0.601 & 0.922& 0 &3\\
rs4844610& 1 &207629207& CR1& 1.84e-09
&0.003016& 0.32 &0.637& 0.799&1 & 3\\
\hline

rs2070116 & 22 & 42606257 & OGFRP1
& 0.1129 & 0.9008 & 0.0413
& 4.51e-9 & 0.0157 & 1 & 2 \\

rs1671152 & 19 & 55526345 & GP6-AS1
& 0.3739 & 0.7508 & 0.701
& 4.55e-7 & 1.09e-13 & 2& 2 \\

rs4803824 & 19 & 45951278 & ERCC1
& 0.9185 & 0.5134 & 0.692
& 2.41e-8 & 0.00047 & 0 & 2 \\
\hline
rs7210399 & 17 & 27780216 & TAOK1
& 0.176 & 0.0596 & 0.806
& 1.96e-6 & 8.24e-9 & 2(LD with rs1967556) & 2 \\

rs1298104 & 16 & 4436445 & TFAP4
& 0.3999 & 0.0642 & 0.906
& 1.34e-8 & 0.0121 & 0& 2 \\

rs6580981 & 12 & 54723028 & COPZ1
& 0.7433 & 0.4933 & 0.654
& 0.00271 & 1.25e-7 & 0& 2 \\

rs79722572 & 12 & 48171222 & SLC48A1
& 0.7946 & 0.0619 & 0.664
& 5.09e-8 & 0.0307 & 0& 2 \\
\hline
rs10822145 & 10 & 63174788 & JMJD1C & 9.95e-01 & 1.72e-01 & 1.19e-01 & 5.47e-17 & 3.85e-13 &  2 (LD with rs7896518) & 2\\
rs113422568 & 10 & 102578181 & SUFU & 4.77e-01 & 3.49e-01 & 6.59e-02 & 1.82e-04 & 5.54e-06 &  0& 2\\
rs61469632 & 9 & 132986603 & GFI1B & 8.95e-01 & 9.53e-02 & 2.28e-01 & 1.69e-11 & 1.74e-03 & 1 & 2\\
rs467369 & 9 & 134040643 & BRD3 & 7.28e-01 & 5.53e-01 & 7.06e-01 & 4.39e-07 & 6.68e-03 & 0&2 \\
rs409801 & 9&4744743 & AK3&0.5085 &0.9985 & 0.165&6.55e-08&0.000377 & 0 &2 \\
\hline

rs4734879 & 8 & 106583124 & ZFPM2
& 0.0189 & 0.6664 & 0.777
& 4.42e-9 & 6.14e-13 & 2& 2 \\

rs342296 & 7 & 106372903 & LINC02577
& 0.0062 & 0.5463 & 0.153
& 2.57e-172 & 1.79e-12 & 2& 2 \\

rs9389269 & 6 & 135427159 & HBS1L
& 0.6391 & 0.0778 & 0.367
& 2.28e-5 & 3.68e-9 & 1& 2 \\

rs9260061 & 6 & 29908491 & LOC105375010
& 0.00352 & 0.00116 & 0.886
& 1.73e-6 & 0.13 &0 & 2 \\

rs10900809 & 5 & 131826322 & IRF1
& 0.1267 & 0.6359 & 0.588
& 5.67e-6 & 4.11e-10 & 2& 2 \\

rs115318824 & 5 & 88296994 & MEF2C-AS1
& 0.1412 & 0.4071 & 0.449
& 1.49e-5 & 0.00558 & 0 & 2 \\

rs1501600 & 3 & 190621019 & GMNC
& 0.0232 & 8.43e-29 & 0.00291
& 0.00953 & 0.101 & 1& 2 \\

rs1354034 & 3 & 56849749 & ARHGEF3
& 0.5251 & 0.353 & 0.193
& 2.23e-6 & 5.24e-14 & 2& 2 \\

rs6733839 & 2 & 127892810 & BIN1
& 1.65e-6 & 1.26e-5 & 0.261
& 0.601 & 0.922 &0 & 2 \\

rs4844610 & 1 & 207802552 & CR1
& 1.84e-9 & 0.00302 & 0.32
& 0.637 & 0.799 &1 & 2 \\

rs750439 & 1 & 156430749 & MIR9-1HG
& 0.7485 & 0.7356 & 0.408
& 5.55e-11 & 0.00567 & 1& 2 \\

\hline
\end{tabular}%
}
\caption{Independent SNPs with LD
threshold of $r^2=0.1$ and at least 500 kb apart, identified by the lfdr-VGM and SQT, with FDR level 0.05 and FWER level 0.05, respectively. The first column gives the RSID of each selected SNP.
The second and third columns give the chromosome number and genomic position,
respectively, and the fourth column gives the nearest gene. The next five
columns report the GWAS $p$-values for A$\beta$, p-tau, NFL, SNAP-25, and
neurogranin, respectively. The final two columns report the results from the
SQT and lfdr-VGM procedure, respectively.
}
\label{tab:seq_lfdr_independent_snps_pen_0}
\end{table}

\end{document}